\documentclass[twocolumn,aps,pre,amsmath,showpacs]{revtex4-1}
\usepackage{graphicx,dcolumn,bm,amssymb,amsmath,ulem,indentfirst,amsthm}
\usepackage[toc,page]{appendix}
\usepackage{color}

\theoremstyle{plain}
\def\be{\begin{equation}}
\def\ee{\end{equation}}

\newtheorem*{theorem*}{Theorem}

\begin{document}

\author{Bingyu Cui$^{1,2}$, Jie Yang$^{3,4}$, Jichao Qiao$^{5}$, Minqiang Jiang$^{3,4}$, Lanhong Dai$^{3,4}$}
\author{Yun-Jiang Wang$^{3,4}$}
\email{yjwang@imech.ac.cn}
\author{Alessio Zaccone$^{1,6}$}
\email{az302@cam.ac.uk}
\affiliation{${}^1$Statistical Physics Group, Department of Chemical
Engineering and Biotechnology, University of Cambridge, CB3
0AS Cambridge, U.K.}
\affiliation{${}^2$Institute for Solid State Physics, The University of Tokyo, 5-1-5 Kashiwanoha, Kashiwa, 277-8581, Japan}
\affiliation{${}^3$State Key Laboratory of Nonlinear Mechanics, Institute of Mechanics, Chinese Academy of Sciences,
Beijing 100190, China}
\affiliation{${}^4$School of Engineering Science, University of Chinese Academy of Sciences, Beijing 101408, China}
\affiliation{${}^5$School of Mechanics, Civil Engineering and Architecture, Northwestern Polytechnical University, Xi'an 710072, China}
\affiliation{${}^6$Cavendish Laboratory, University of Cambridge, JJ Thomson
Avenue, CB3 0HE Cambridge,
U.K.}

\begin{abstract}
An atomic-scale theory of the viscoelastic response of metallic glasses is derived from first principles, using a Zwanzig-Caldeira-Leggett system-bath Hamiltonian as a starting point within the framework of nonaffine linear response to mechanical deformation.
This approach provides a Generalized-Langevin-Equation (GLE) as the average equation of motion for an atom or ion in the material, from which non-Markovian nonaffine viscoelastic moduli are extracted. These can be evaluated using the vibrational density of states (DOS) as input, where the boson peak plays a prominent role for the mechanics.
To compare with experimental data of binary ZrCu alloys, numerical DOS was obtained from simulations of this system, which take also electronic degrees of freedom into account via the embedded atom method (EAM) for the interatomic potential. It is shown that the viscoelastic $\alpha$-relaxation, including the $\alpha$-wing asymmetry in the loss modulus, can be very well described by the theory if the memory kernel (the non-Markovian friction) in the GLE is taken to be a stretched-exponential decaying function of time. This finding directly implies strong memory effects in the atomic-scale dynamics, and suggests that the $\alpha$-relaxation time is related to the characteristic time-scale over which atoms retain memory of their previous collision history. This memory time grows dramatically below the glass transition.
\end{abstract}

\title{Atomic theory of viscoelastic response and memory effects in metallic glasses}
\maketitle

\section{Introduction}
The mechanism by which supercooled liquids undergo a liquid-solid transition at or around the glass transition temperature, $T_{g}$, has remained elusive~\cite{Donth,Ngai,Goetze}.
The $\alpha$-relaxation process describes the slow decay of density correlations and is typically related to the intermediate scattering function, although it can be also  observed in the mechanical
relaxation, as well as in the dielectric
response~\cite{Richert}.
Within the energy landscape picture, the $\alpha$-relaxation can be interpreted as the transition
of the system from one meta-basin to another, by means of a collective thermally
activated jump over a large energy barrier, a process that, for high-dimensional systems, can be well described by replica symmetry-breaking and related approaches~\cite{Zamponi,Mezard,Yoshino}.
While the calorimetric glass transition may be quite smooth, the vanishing of the low-frequency shear modulus near $T_{g}$ can be, instead, very dramatic, with a sudden drop by orders of magnitude~\cite{Crespo} that can be related to marginal stability~\cite{ZacconePRL}. 

Compared with traditional disordered materials, metallic glasses (MGs) exhibit extraordinary physical properties, in terms of their ability to sustain large loads prior to yielding and their ductility~\cite{Spaepen}. However, although previous atomic scales theories based on defect physics and lattice dynamics have provided a good understanding of mechanical relaxation and internal friction in crystalline metals~\cite{Nabarro}, unravelling from the same microscopic scale the relation between viscoelasticity and dynamical heterogeneity for metallic glasses has been a long-term challenge.

With the advent of MGs as the next-generation metallic materials for technological applications, extensive experimental investigations like stress relaxation technique have brought
a wealth of observations about the viscoelasticty and anelasticity of these materials. A lot of research has been taken on the stress relaxation of various metallic glasses, which claims that localized plastic flow could be activated during viscoelastic and plastic deformation~\cite{Wang, Qiao}. The whole relaxation spectrum of viscoelastic materials is usually fitted by the Kohlrausch (stretched exponential) function which is simply an empirical model, hence does not arise from any physical mechanism. 

The most used and successful microscopic framework that has been applied to the atomic and molecular dynamics and relaxation of supercooled liquids above $T_g$ is Mode-Coupling-Theory (MCT)~\cite{Goetze-book,Voigtmann}. Other theories have focused on the mesoscopic-level description of nonlinear deformation such as the Shear-Transformation-Zone (STZ)~\cite{Argon,Langer,Falk}. A recent theory based on coherent-potential approximation and on the continuum assumption of heterogeneously fluctuating modulus has achieved success in the comparison with experimental data of linear dynamic moduli of metallic glasses~\cite{Mazzone}, but does not provide microscopic atomic-scale insights given its continuum macroscopic character and does not account for electronic effects.

The main limitations for developing an atomic-scale theory of viscoelasticity and of the dynamic mechanical response of MGs are as follows:
(i) the atomic-scale dynamics of glasses under deformation is strongly nonaffine~\cite{Zaccone2011,Hufnagel}, meaning that additional displacements on top of the affine displacements prescribed by the strain tensor, are required to relax quenched neighbouring forces caused by the lack of centrosymmetry of the disordered lattice~\cite{Milkus}; (ii) the vibrational density of states (DOS) which governs the atomic-scale dynamics is rich in low-energy soft modes (boson peak) whose physical origin has been elusive~\cite{Loffler,Albe}, and only recently have been traced back to mesoscopic phonon scattering processes and the Ioffe-Regel crossover~\cite{Shintani}, which are also in relation to the lack of centrosymmetry; (iii) there is currently no established understanding for the atomic-scale internal friction, which is crucial to deriving viscoelastic sum-rules, and is associated with memory effects which are known to be important for metallic glass~\cite{Wei-Hua_PRL}; (iv) the interatomic interaction is strongly non-local, also due to the role of delocalized electrons which affect the interatomic interaction (see Appendix C).

Here we provide an answer to all these issues in a unifying way, by deriving a nonaffine atomic-scale theory of viscoelastic response and relaxation of metallic glasses, in a bottom up way starting from a microscopic Hamiltonian. We use the Zwanzig-Caldeira-Leggett (ZCL) system-bath Hamiltonian to derive an average equation of motion for a tagged atom (or ion), which turns out to be a Generalized Langevin Equation (GLE), with a non-Markovian atomic-scale friction (memory kernel). The latter memory kernel arises from integrating out the fast degrees of freedom of the atomic motion~\cite{Zwanzig,Weiss}. Although it is currently not possible to specify the functional form of the time-dependence of the friction within ZCL models, a stretched-exponential form for the microscopic friction in supercooled liquids was derived by Sjoegren and Sjoelander based on many-body kinetic theory~\cite{Sjoegren}.

In order to test the theory we use stress-relaxation experiments on \textrm{Cu}$_{50}$\textrm{Zr}$_{50}$ glassy system.
Furthermore, the vibrational DOS is needed as input to calculate the viscoelastic response. To this aim, we used numerical simulations of the same metallic glasses which take also electronic structure effects into account at the level of the embedded atom method (EAM).\\

\section{Experiments}
Thanks to the thermal stability of CuZr- based metallic glass. MG ribbons made up of \textrm{Cu}$_{50}$\textrm{Zr}$_{50}$ with length over 7 mm were processed by the melt-spinning technique in an inert argon atmosphere. Differential scanning calorimetry (DSC) was used to determine the thermal properties of the samples that has a glass transition temperature $T_{g}$ at 670 K at a heating rate of 20 K/min. The tensile stress relaxation experiments were performed with a TA Q800 dynamic mechanical analyzer (DMA). To eliminate any influences from initial states, the MG ribbons were heated above $T_{g}$ before the measurements. The tensile stress relaxation, carried out at a constant strain of $0.4\%$ was loaded on the model alloy for 24 hours after an initial 3 minutes equilibrium. The resultant stress relaxation in a form of time dependency that is fitted by the Kohlrausch function $\sigma(t)=\sigma_0\exp[-(t/\tau)^{\beta}]$ with $\sigma_0$ being stress relaxation at $t=0$, is given in Fig. 1, under three different temperatures $T_g $(670 K), 0.9 $T_g $(603 K) and 0.8 $T_g$ (536 K). Note that, we have roughly estimated $\sigma_{\infty}$, which is $\sigma(t)$ at $t=\infty$, to be zero for the three temperatures.

Note that in Fig. 1 the fitting is excellent apart from deviations which are due to processes other than the $\alpha$-relaxation (e.g. other long-time or low-frequency relaxation processes). In this work we want to focus on a theory of $\alpha$-relaxation and its associated viscoelastic response, without considering other processes. In the following, we will use the fitted Kohlrausch function to obtain the dynamic moduli $E'$ and $E''$ in the frequency domain (see Appendix D). In this way, we will be targeting the $\alpha$-relaxation only, and consistently focus our attention on the comparison between our theory (for $\alpha$-relaxation) and data extracted from experiments where effects other than $\alpha$-relaxation have been removed.

\section{MD simulations with EAM potentials}
In molecular dynamics (MD) simulations, we utilized the Finnis-Sinclair type EAM potentials optimized for realistic amorphous Cu-Zr structures~\cite{Mendelev}. Seven independent \textrm{Cu}$_{50}$\textrm{Zr}$_{50}$ MG models were obtained by quenching the system at cooling rate $10^{10}$ K/s from a liquid state equilibrated at 2000 K with different initial position and velocity distribution. Each model was composed of 8192 atoms and external pressure was held at zero during the quenching process using a Parrinello-Rahman barostat~\cite{Parrinello}. Periodic boundary conditions were imposed automatically. The resulting vibrational DOSs averaged from seven independent glassy models are shown in Fig. 2. It can be easily seen that the eigenfrequency spectrum is not sensitive to temperature.\\

\section{Theory}
In condensed matter physics, the ZCL system-bath model is widely applied to low-temperature quantum physics problems, especially in quantum tunnelling in superconductors and in chemical reaction rate theory. Aiming at deriving a suitable equation of motion for a tagged atom (or ion) in a metallic glass, we extend this approach to atomic dynamics in disordered materials by taking into account the disordered environment as well as the dissipation. In the construction of this approach, it is well known that one cannot consider anharmonicity~\cite{Weiss}. However, anharmonicity is indirectly taken into account in our framework through both the vibrational DOS and the emergent friction kernel as shown below.\\

In the ZCL approach, the Hamiltonian of a tagged atom coupled to all the very many other atoms in the material (treated as harmonic oscillators) is given by~\cite{Zwanzig}
\begin{equation}
H=\frac{P^2}{2m}+V(Q)+\frac{1}{2}\sum_{\alpha=1}^N\left[\frac{p_{\alpha}^2}{m_{\alpha}}+m_{\alpha}\omega_{\alpha}^2\left(X_{\alpha}
-\frac{F_{\alpha}(Q)}{m_{\alpha}\omega_{\alpha}^2}\right)^{2}\right]
\end{equation}
where the first two terms are the Hamiltonian of tagged particle with (effective) mass $m$, while $\frac{1}{2}\sum_{\alpha=1}^N(\frac{p_{\alpha}^2}{m_{\alpha}}+m_{\alpha}\omega_{\alpha}^2 X_{\alpha}^2)$ is the Hamiltonian of the bath of harmonic oscillators coupled to the tagged particle with linear coupling function $F_{\alpha}(Q)=c_{\alpha}Q$ where $c_{\alpha}$ are the coupling strength coefficients which are different for all the different atoms the tagged atom is interacting with (e.g. $c_{\alpha}$ is expected to be large for nearby atoms and small for atoms far away in the material).
This configuration gives rise to a second-order inhomogeneous differential equation for the position of the $\alpha$-th oscillator of the bath, whose solution leads to the following GLE:
\[m\ddot{Q}=-V'(Q)-\int_{-\infty}^t\frac{\nu(t-t')}{m}\frac{dQ}{dt'}dt' + F_{p}(t).\]

As is standard for normal mode analysis, we introduce the rescaled tagged-particle displacement $q=Q\sqrt{m}$ in the Hamiltonian, such that the resulting equation of motion, using mass-rescaled coordinates, becomes
\begin{equation}
\ddot{q}=-V'(q)-\int_{-\infty}^t \nu(t-t')\frac{dq}{dt'}dt' + F_{p}(t).
\end{equation}
Upon focusing on the athermal limit of the dynamics for $T<T_{g}$, the noise term $F_{p}(t)$ can be ignored which amounts to assuming low thermal noise and frozen-in atomic
positions, which is a meaningful approximation below $T_g$. Also, for dynamical response to an oscillatory strain one can average the dynamical equation over many cycles, which amounts to a time-average. Since the noise $F_{p}$ has zero-mean~\cite{Zwanzig}, an average over many cycles could be effectively similar to an ensemble average thus leaving $\langle F_{p} \rangle=0$ in the above equation. Since the system is non-ergodic below $T_{g}$, nothing guarantees that this is true a priori, but there is initial evidence that this approximation might be reasonable in the linear regime where the response converges to a reproducible noise-free average stress~\cite{Rodney}.

As shown in Ref. \cite{Zwanzig},
the friction coefficient $\nu$ arises from the long-range coupling between atoms in the ZCL model, which effectively takes care of long-range and many-body anharmonic tails of interatomic interaction (see Appendix A for further discussion about $\nu$). In our theory, the effect of $T$ is taken care of by the DOS and also the parameters of the memory kernel will turn out to be $T$-dependent, as shown below.

Upon applying a deformation described by the strain tensor $\underline{\underline{\eta}}$, the nonaffine dynamics of a tagged particle $i$ interacting with other atoms satisfies the following equation for the displacement $\{\underline{x}_i(t)=\underline{\mathring{q}}_i(t)-\underline{\mathring{q}}_i\}$ around a known rest frame $\underline{\mathring{q}}_i$ (see Appendix B for details of derivation):
\begin{equation}
\frac{d^2\underline{x}_i}{dt^2}+\int_{-\infty}^{t}\nu(t-t')\frac{d\underline{x}_i}{dt'}dt'+\underline{\underline{H}}_{ij}\underline{x}_j=\underline{\Xi}_{i,xx}\eta_{xx},
\end{equation}
which can be solved by performing Fourier transformation followed by normal mode decomposition that decomposes the 3N-vector $\tilde{\underline{x}}$, that contains positions of all atoms, into normal modes $\tilde{\underline{x}}=\hat{\tilde{x}}_p(\omega)\underline{\phi}^p$ ($p$ is the index labeling normal modes).  Note that we specialize on time-dependent uniaxial strain $\eta_{xx}(t)$. For this case, the vector $\underline{\Xi}_{i,xx}$ represents the force per unit strain acting on atom $i$ due to the motion of its nearest-neighbors which are moving towards their respective affine positions (see e.g.~\cite{Lemaitre} for a more detailed discussion) and in our case also includes electronic effects empirically via the EAM potential (see Appendix C).\\

From now on we drop all $i$ and $j$ indices, and all matrices and vectors are meant to be $3N \times 3N$ and $3N$-dimensional, respectively.
After taking Fourier transformation, we have
\[-\omega^2\tilde{\underline{x}}+i\tilde{\nu}(\omega)\omega\tilde{\underline{x}}
+\underline{\underline{H}}~\underline{\tilde{x}}
=\underline{\Xi}_{xx}\tilde{\eta}_{xx}.
\]

Next, we take normal mode decomposition. This is equivalent to diagonalise the matrices $\underline{\underline{H}}$. The $3N\times3N$ matrix
$\underline{\underline{H}}$ can be decomposed as $\underline{\underline{H}}=\underline{\underline{\Phi}}~\underline{\underline{D}}~\underline{\underline{\Phi}}^{-1}=\underline{\underline{\Phi}}~\underline{\underline{D}}~\underline{\underline{\Phi}}^T$ where $\underline{\underline{D}}$ is a diagonal matrix filled with the eigenvalues of $\underline{\underline{H}}$, that is, in components, $D_{pp}=\omega_p^2$. Further, the matrix $\underline{\underline{\Phi}}$ consists of the eigenvectors  $\underline{\phi}_i$ of the Hessian, i.e. $\underline{\underline{\Phi}}=(\underline{\phi}^1,...,\underline{\phi}^p,...,\underline{\phi}^{3N})$, and is an orthogonal matrix.  First, we left-multiply both sides with the matrix $\underline{\underline{\Phi}}^{-1}=\underline{\underline{\Phi}}^T$,
\[-\omega^2(\underline{\underline{\Phi}}^{T}\tilde{\underline{x}})+i\tilde{\nu}(\omega)\omega\underline{\underline{\Phi}}^{T}\tilde{\underline{x}}
+\underline{\underline{D}}~(\underline{\underline{\Phi}}^{T}\underline{\tilde{x}})
=\underline{\underline{\Phi}}^{T}\underline{\Xi}_{xx}\tilde{\eta}_{xx},
\]
where we used the fact that $\underline{\underline{D}}$ is diagonal.
From the definition of $\underline{\underline{\Phi}}^{T}$, we have $\underline{\underline{\Phi}}^{T}\underline{\tilde{x}}=(\underline{\tilde{x}}\cdot\underline{\phi}^1,...,\underline{\tilde{x}}\cdot\underline{\phi}^p
,...,\underline{\tilde{x}}\cdot\underline{\phi}^{3N})^{T}$. That is, if we rewrite the above equation as a system of $3N$ linear equations, the equation for the p-th mode reads:
\[-\omega^2\underline{\tilde{x}} \cdot \underline{\phi}^{p}+i\omega\tilde{\nu}(\omega)\underline{\tilde{x}} \cdot \underline{\phi}^{p}
+\omega_p^2\underline{\tilde{x}} \cdot \underline{\phi}^{p}
=\underline{\Xi}_{xx} \cdot \underline{\phi}^{p}\tilde{\eta}_{xx}.\]
\\

We recall the definition of normal modes as $\tilde{\underline{x}}(\omega)=\hat{\tilde{x}}_p(\omega)\underline{\phi}^p$ and $\hat{\tilde{\underline{x}}}_p(\omega)=\tilde{\underline{x}}(\omega)\cdot \underline{\phi}^p$ where hat denotes the coefficient of the projected quantity. Thus, we obtain
\[-\omega^2\hat{\tilde{x}}_p(\omega)+i\tilde{\nu}(\omega)\omega\hat{\tilde{x}}_p(\omega)
+\omega_p^2\hat{\tilde{x}}_p(\omega)
=\hat{\tilde{x}}_p(\omega)\tilde{\eta}_{xx},\]
from which an explicit expression for $\hat{\tilde{x}}_p(\omega)$ can easily be found.

It was shown in previous work~\cite{Lemaitre} that $\hat{\Xi}_{xx}$ is self-averaging, and one can introduce the smooth correlator function $\Gamma_{xxxx}(\omega)=\langle\hat{\Xi}_{p,xx}\hat{\Xi}_{p,xx}\rangle_{p\in\{\omega,\omega+\delta\omega\}}$ on frequency shells.
Following the general procedure of Ref.~\cite{Lemaitre} to find the oscillatory stress for a dynamic nonaffine deformation, the stress is obtained to first order in strain amplitude as a function of $\omega$, as
\begin{align}
\tilde{\sigma}_{xx}(\omega)&=E_A\tilde{\eta}(\omega)-\frac{1}{V}\sum_p\hat{\Xi}_{p,xx}\hat{\tilde{x}}_p(\omega) \notag\\
&=E_A\tilde{\eta}(\omega)+\frac{1}{V}\sum_p\frac{\hat{\Xi}_{p,xx}\hat{\Xi}_{p,xx}}{\omega^2-\omega_p^2
-i\tilde{\nu}(\omega)\omega}\tilde{\eta}(\omega)\notag\\
&=E_{xxxx}(\omega)\tilde{\eta}(\omega).
\end{align}

In the thermodynamic limit with continuous spectrum, we replace the discrete sum over $3N$ degrees of freedom with an integral over eigenfrequencies up the Debye frequency $\omega_{D}$, and we thus obtain the complex Young's modulus as
\begin{equation}
E^*(\omega)=E_A-3\rho\int_0^{\omega_{D}}\frac{D(\omega_p)\Gamma(\omega_p)}{\omega_p^2-\omega^2+i\tilde{\nu}(\omega)\omega}d\omega_p
\end{equation}
where we have dropped the Cartesian indices for convenience, since we are specializing on uni-axial extension, and $\rho=N/V$ denotes the atomic density of the solid.\\
This is a crucial result of this paper, derived here for the first time. It differs from a previous result obtained in Ref.~\cite{Lemaitre} because the friction coefficient is non-Markovian, hence frequency-dependent, whereas in Ref.~\cite{Lemaitre} it is just a constant, corresponding to Markovian dynamics. This will turn out to be a fundamental difference, because as we show below, metallic glass data cannot be described by a friction coefficient which is constant with frequency. Furthermore, this result is derived here from a microscopic Hamiltonian.

In the numerical simulations, the DOS is actually not a continuous function, but discrete. Thus, in Eq.(5) we replace the DOS with its spectral representation given by a sum of delta-functions. Since under each temperature, we have seven simulated samples with different configurations for position and velocity to calculate the DOS, we take the same fitting parameters for each sample and found that they all generate the same results. Hence, in Fig. 3 and 4, we simply show the results from one out of these seven simulated systems.
The DOS is calculated by diagonalizing the Hessian matrix for the interaction energy of an atom in CuZr alloys in mass-rescaled coordinates, which is also used to calculate the $\underline{\Xi}_i$ vectors and hence $\Gamma(\omega_p)$. Analytical expressions for the Hessian and for $\underline{\Xi}_i$ as a function of the EAM interaction have been derived in the Appendix C.

We then rewrite $E^*(\omega)$ as a sum over a discrete distribution of $\omega_p$ from the MD simulation of DOS, $E^*(\omega)=E'(\omega)+iE''(\omega)$:
\begin{align}
E'(\omega)=E_A-A\sum_p\frac{\Gamma(\omega_p)(\omega_p^2-\omega^2+\tilde{\nu}_2\omega)}{(\omega_p^2-\omega^2+\tilde{\nu}_2\omega)^2+(\omega\tilde{\nu}_1)^2}\\
E''(\omega)=B\sum_p\frac{\Gamma(\omega_p)(\omega\tilde{\nu}_1)}{(\omega_p^2-\omega^2+\tilde{\nu}_2\omega)^2+(\omega\tilde{\nu}_1)^2}
\end{align}
where $E_A$, $A$, and $B$ are rescaling constants to be calibrated in the comparison. $\tilde{\nu}_1$ and $\tilde{\nu}_2$ are the real and (minus) imaginary parts of $\tilde{\nu}(\omega)$ that is the Fourier transform of $\nu(t)$, $\tilde{\nu}(\omega)=\tilde{\nu}_1(\omega)-i\tilde{\nu}_2(\omega)$. We have chosen the ansatz of $\nu(t)=\nu_0e^{-rt^b}$ motivated by previous theoretical work ~\cite{Sjoegren}, where $b = 0.3$ was found to work well for molecular glasses in Ref.~\cite{Cui}. Here, a larger value of $b$ appears appropriate for metallic glass~\cite{Ruta}. \\

\begin{figure}
\begin{center}
\includegraphics[height=5.4cm,width=8.3cm]{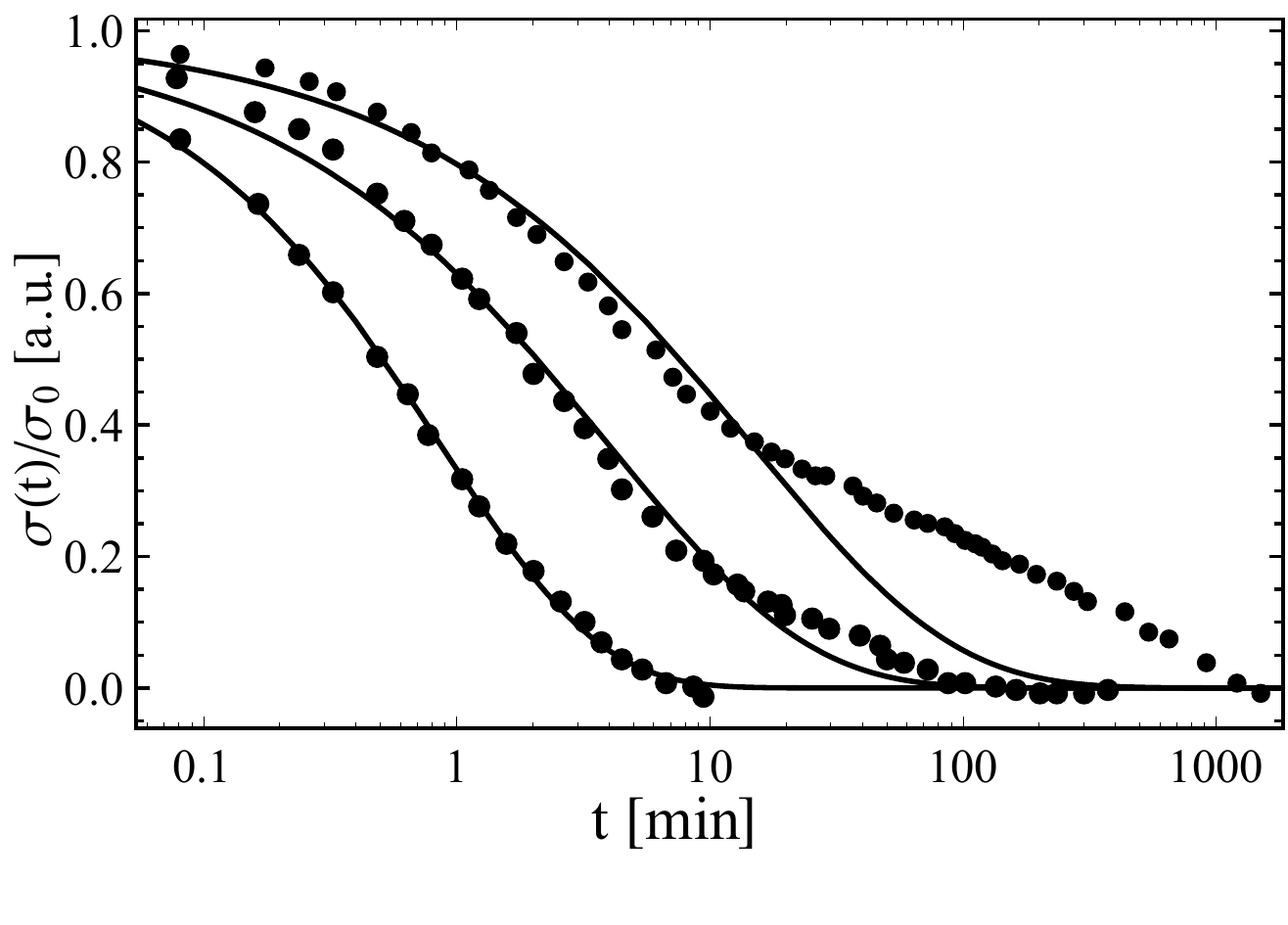}
\caption{Kohlrausch empirical fits (solid lines) of experimental data (symbols). Top to bottom corresponds to temperatures in the following order: 536 K, 603 K, 670 K($T_g$). Solid curves are Kohlrausch $\sigma(t)\sim\exp{-(t/\tau)^\beta}$ empirical fittings used to calibrate results, where the two parameters $\beta$ and $\tau$ were chosen to be 0.69, 0.87 (mins); 0.55, 4.03 (mins); 0.55, 14.87 (mins) for $T_g$, 0.9 $T_g$ and 0.8 $T_g$ respectively.}
\end{center}
\end{figure}

Apart from ZCL Hamiltonian, the Nose-Hoover method also provides a route towards estimating the time-dependent non-Markovian friction~\cite{Hoover}. After carrying simulation in the canonical ensemble below $T_g$, one obtains a simple-exponential decay of the friction coefficient, with which however one cannot reproduce the experimentally-measured curves of $E'$ and $E''$, over any interval in frequency. This problem might be due to the limitations of using the Nose-Hoover method for nonequilibrium systems.

In general, the determination of the memory kernel is an open problem for which several approaches have been proposed very recently, most of which have been tested only on model systems so far~\cite{Karniadakis1, Karniadakis2, Schmidt, Schilling, Izvekov}. In future work, our proposed framework can be combined with projection-operator methods~\cite{Karniadakis1,Schilling} to derive the memory kernel used here from first principles.

\section{Results and discussion}
Before presenting a comparison between our theory with the empirical best-fitting Kohlrausch stretched-exponential relaxation fitting of experimental
data on \textrm{Cu}$_{50}$\textrm{Zr}$_{50}$, we firstly convert the linear response of the material to applied stress from time-dependent compliance to the frequency-dependent dynamic moduli, for a uniaxial strain of amplitude $\epsilon_0$:
\begin{align}
E'(\omega)&=\frac{\sigma_{\infty}}{\epsilon_0}+\frac{\sigma_0\omega}{\epsilon_0}\int_0^{\infty}e^{-(t/\tau)^{\beta}}\sin{\omega t}dt\\
E''(\omega)&=\frac{\sigma_0\omega}{\epsilon_0}\int_0^{\infty}e^{-(t/\tau)^{\beta}}\cos{\omega t}dt.
\end{align}
A detailed derivation of this result is reported in Appendix D.

\begin{figure}
\begin{center}
\includegraphics[height=5.4cm,width=8.5cm]{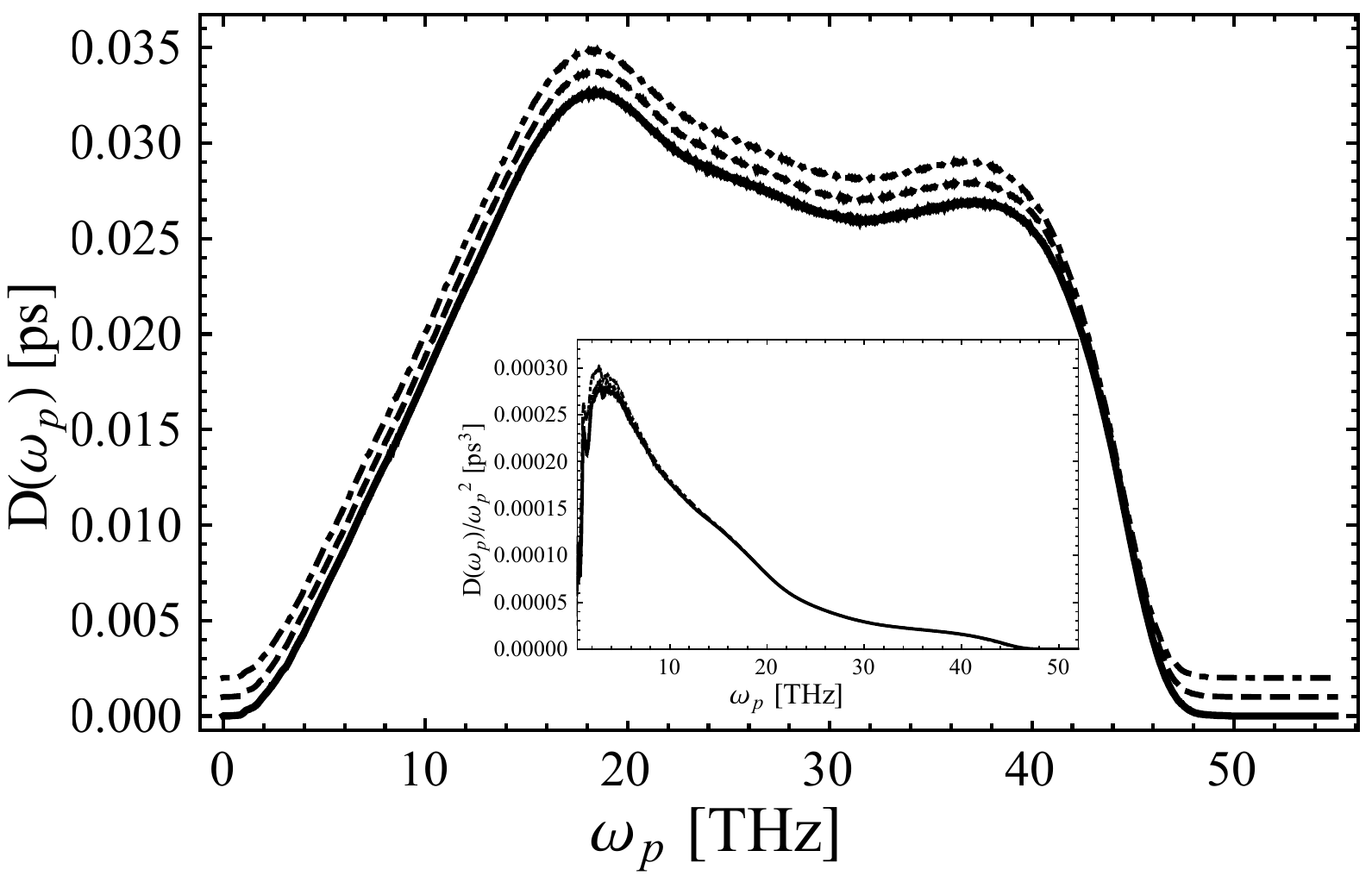}
\caption{Vibrational density of states (DOS) from simulated \textrm{Cu}$_{50}$\textrm{Zr}$_{50}$ system. Solid, dashed and dotted lines correspond to DOS at 670 K, 603 K and 536 K, respectively. The curves have been lifted upward in order to be distinguishable for the reader. The inset shows the DOS normalized by the Debye law $\omega_{p}^{2}$ which shows clear evidence of a strong boson peak.}
\end{center}
\end{figure}

\begin{figure}
\begin{center}
\includegraphics[height=5.8cm,width=8.5cm]{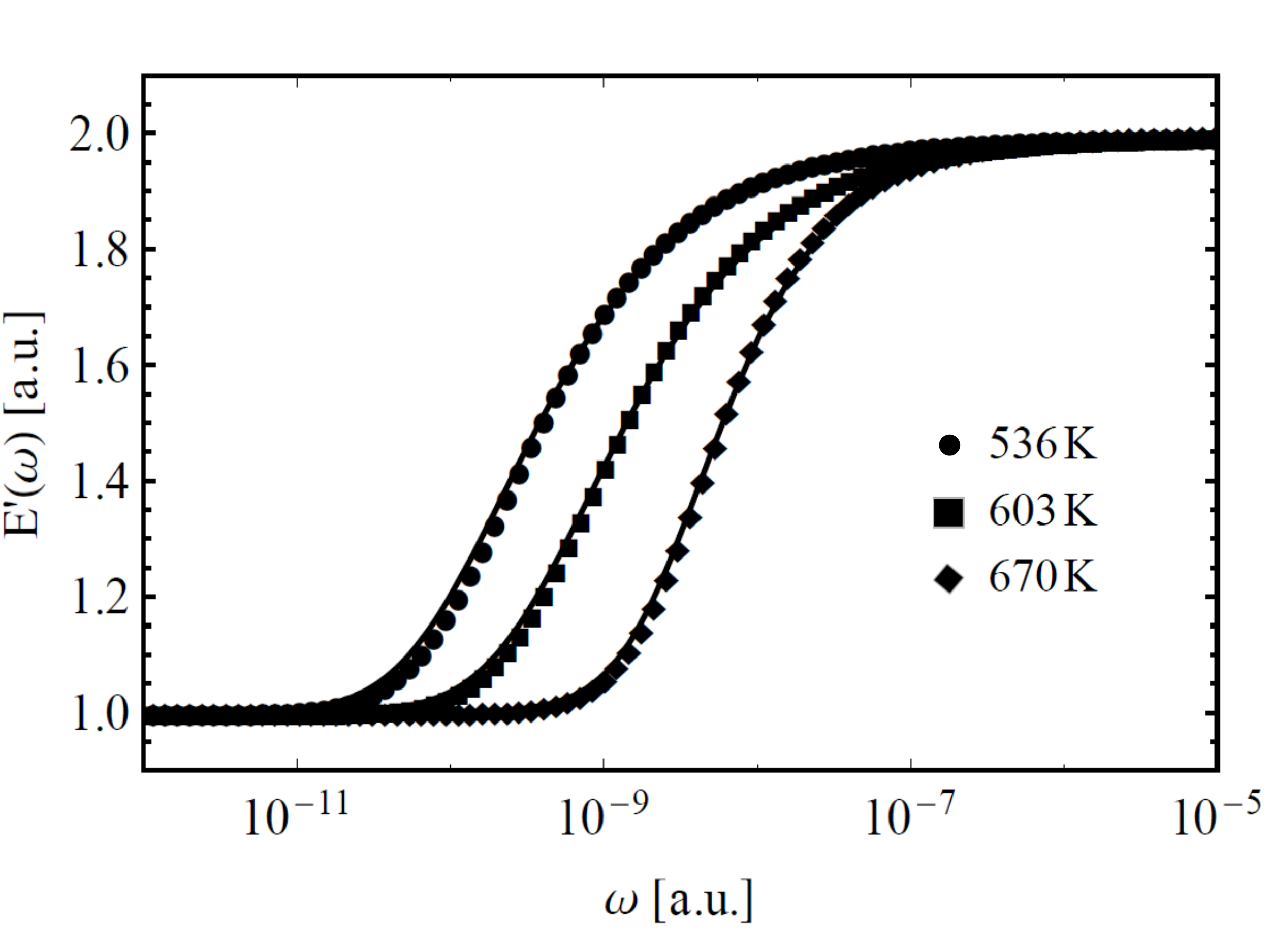}
\caption{Real part of the complex viscoelastic modulus. From right to left solid lines represent $E'$ for $T_g$, 0.9 $T_g$ and 0.8 $T_g$ respectively, from the Kohlrausch best fitting of our experimental data. Symbols are calculated based on our theory. For $T_g$, 0.9 $T_g$ and 0.8 $T_g$, $b$ was chosen to be 0.72, 0.58 and 0.58; $r$ was taken to be $1.2 \times 10^{-6}$, $7 \times 10^{-6}$ and $3.4 \times 10^{-6}$. $\nu(0)$=0.137 is same for all temperatures. Rescaling constants have been taken to adjust the height. }
\end{center}
\end{figure}

\begin{figure}
\begin{center}
\includegraphics[height=5.8cm,width=8.6cm]{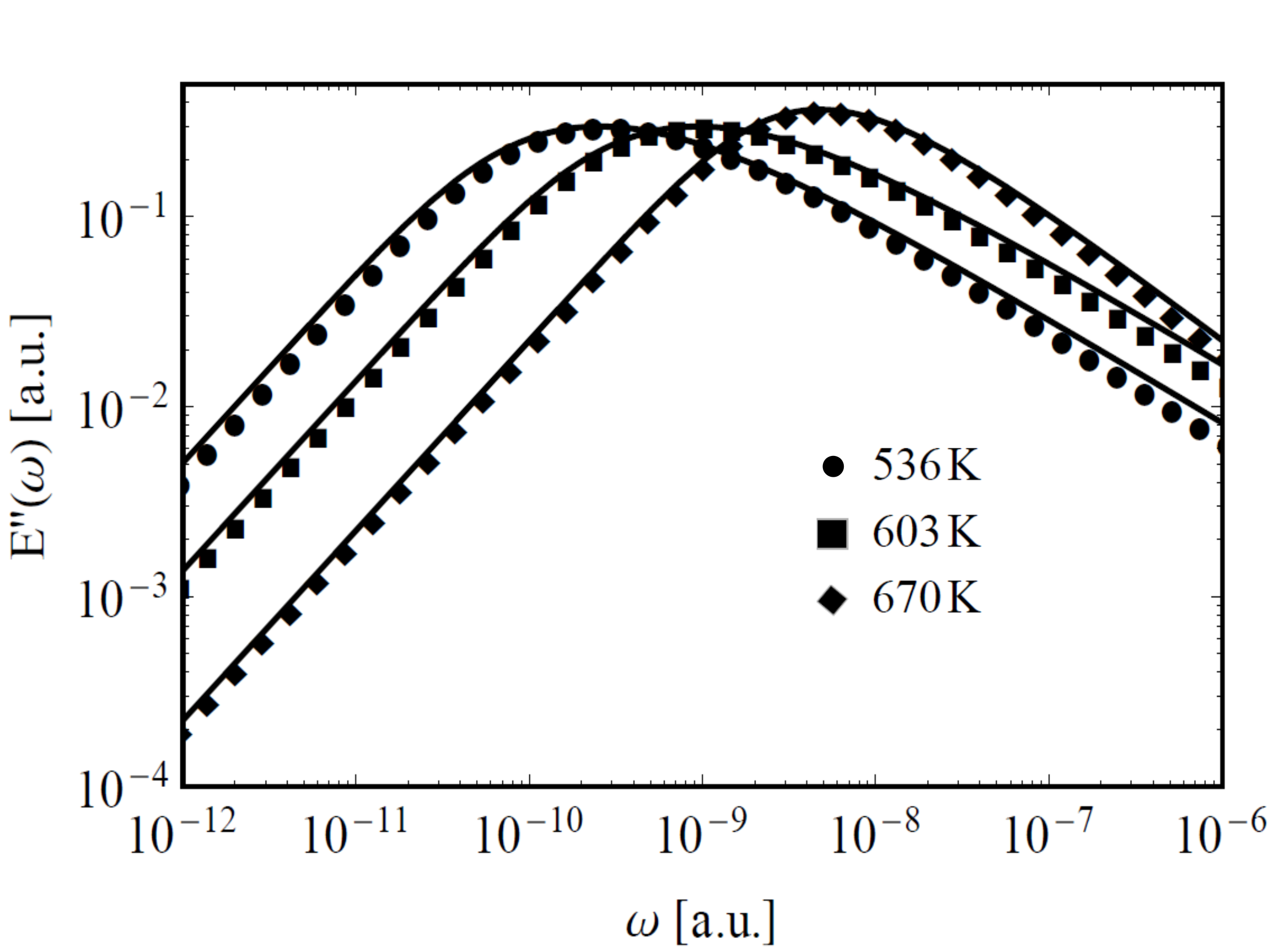}
\caption{Imaginary part of the complex viscoelastic modulus. From right to left solid lines represent $E''$ for $T_g$, 0.9 $T_g$ and 0.8 $T_g$ respectively, from empirical Kohlrausch fittings of the experimental data. Symbols are calculated from our theory. For $T_g$, 0.9 $T_g$ and 0.8 $T_g$, $b$ in the memory-kernel of our theory was chosen to be 0.72, 0.58 and 0.58; $r$ was taken to be $1.2 \times 10^{-6}, 7\times10^{-6}$ and $3.4 \times10^{-6}$. $\nu(0)$=0.137 is same for all temperatures. Rescaling constants have been taken to adjust the height.}
\end{center}
\end{figure}

In Fig. 3 we plotted the comparisons for $E'(\omega)$ at $T_g=670$ K, i.e.
exactly at $T_{g}$, from Eq. (6) and Eq. (8).
In this case, it is clear that our theoretical model is in excellent agreement with the transformed experimental data, and is also very close to the Kohlrausch function. This shows how crucial soft modes are, as well as the memory effects embodied in the non-Markovian friction, for the understanding of the viscoelastic response and of $\alpha$-relaxation below the glass transition.
In Fig. 4, we present fittings of the loss modulus, $E''(\omega)$. Also in
this case, it is seen that our theory, given by Eq. (7), provides an excellent description of the experimental data. Note that, for clarity of presentation, we have changed the unit of time to shift curves horizontally. This means we have arbitrary units on abscissa and ordinate.

Remarkably, our theoretical model provides the long-sought crucial and direct connection between the excess of low-energy (boson-peak) modes of the DOS at $T_{g}$, the memory effects in the dynamics, and the corresponding features of the response such as the $\alpha$-wing asymmetry in $E''(\omega)$.
It is in fact impossible to achieve a fitting of the data using a Debye model for the DOS which has no excess of soft modes.

Even more crucially, in contrast with previous approaches, our theory shows that memory effects are as important as the boson peak modes in order to describe the experimental data. We have indeed checked that using a constant (Markovian) friction $\nu=\textrm{const}$, or even a simple-exponential time-dependence for $\nu(t)$, it is not possible to describe the experimental data. Only a stretched-exponential form of $\nu(t)$ with a value of the stretching exponent in the range $0.58 - 0.72$, which decreases upon decreasing $T$ further down from $T_g$, allows us to describe the data. Since $\nu$ in our theory physically represents the spectrum of dynamic coupling between an atom and all other atoms in the material, this result implies that every atom is long-ranged coupled to many other atoms beyond the nearest-neighbour shell, which is the result of the anharmonicity of the interaction and of the non-locality of the electronic contributions to the interatomic interaction.

Also, our theoretical analysis shows that the time-scale over which atoms retain memory of their previous collision history, $\tau_{m}\equiv r^{(-1/b)}$ in our model, also increases upon decreasing the temperature, by more than a factor two overall, even though this increase appears to be somewhat non-monotonic, from $\tau_m\approx 1.67\times 10^{8}$ at $T=T_g$, to $\tau_m\approx 7.72\times 10^{8}$ at $T=0.9 T_g$, to $\tau_m\approx 2.68\times 10^{9}$ at $T=0.8 T_g$.\\

\section{Conclusion}
We have developed a dissipative nonaffine lattice dynamics of metallic glass, in a bottom-up approach starting all the way from a microscopic Hamiltonian for the motion of a tagged atom
coupled to all other atoms in the material.
The theory leads to a Generalized Langevin Equation that we use in combination with nonaffine dynamics to derive the dynamic viscoelastic moduli $E'(\omega)$ and $E''(\omega)$ which are functions of the vibrational DOS and of the emergent non-Markovian atomic-scale friction coefficient (memory kernel) that embodies the long-range coupling between atoms.

The predictions of our theory compare very well with experimental data for uniaxial viscoelastic response of CuZr metallic glasses, using the DOS from MD simulations of the same system.
Importantly, no agreement can be found using either a DOS that does not feature excess of boson-peak modes at low frequency, or using a time-dependence of the non-Markovian friction in the equation of motion which differs from a stretched-exponential function. This finding indicates strong memory effects at the atomic level, possibly due to the non-local electronic component of interaction. It is also shown that the $\alpha$-wing asymmetry in $E''$ grows upon decreasing the temperature below $T_g$, which is linked to the growth of the characteristic time-scale of memory effect in our model, $\tau_m=r^{-(1/b)}$ in our analysis above. Hence, a link exists between the $\alpha$  time and the characteristic time-scale over which atoms retain memory of their previous collision history.

Hence, this analysis establishes that, in order to explain the mechanical $\alpha$-relaxation and the $\alpha$-wing asymmetry in metallic glass, an excess of soft vibrational modes as well as strong memory effects in the dynamics due to non-local electronic coupling between many atoms, are necessary ingredients that cannot be neglected. Furthermore, our approach is directly applicable to a variety of glassy and partly-ordered systems that feature a boson peak, hence not only metallic glasses but also polymer glasses~\cite{Sokolov}, silica glasses~\cite{Barrat} and even quartz~\cite{Chumakov}, by suitably extending the theory to include bond-bending interactions, needed to describe covalent bonds. Hence, this framework opens up the way for a truly atomic-level predictive and quantitative description of mechanical response and relaxation in disordered materials.

\begin{acknowledgements}
Many useful discussions with W. Goetze, E. M. Terentjev, R. Milkus, A. Tanguy and D. Rodney are gratefully acknowledged. Y.J.W acknowledges the financial
support from NSFC (Nos. 11402269, 11672299, 11472287), and the Strategic Priority Research Program of the Chinese Academy of Sciences (No. XDB22040302).
\end{acknowledgements}

\appendix
\section{The memory kernel for the microscopic friction}
The ZCL Hamiltonian does not put any constraint on the form of the memory function $\nu(t)$, which can take any form depending on the values of the coefficients $c_{\alpha}$~\cite{Zwanzig}.
Hence, a shortcoming of ZCL-type models is that, in general, the spectrum of coupling constants 
$\{c_{\alpha}\}$ is not accessible from theory alone.

However, for a supercooled liquid, a  relationship between the time-dependent friction, which is dominated by slow collective dynamics, and the intermediate structure factor has been famously derived within kinetic theory (Boltzmann equation) by Sjoegren and Sjoelander ~\cite{Sjoegren} (see also Ref.\cite{Bagchi}):
\begin{equation}
\nu(t)=\frac{\rho k_{B}T}{6\pi^2 m}\int_{0}^{\infty}dk k^{4} F_{s}(k,t)[c(k)]^{2} F(k,t)
\end{equation}
where $c(k)$ is the direct correlation function of liquid-state theory, $F_{s}(k,t)$ is the self-part of the intermediate scattering function and $F(k,t)$ is the intermediate scattering function~\cite{Sjoegren}. All of these quantities are functions of the wave-vector $k$ and the integral over $k$ leaves a time-dependence of
$\nu(t)$ which is exclusively given by the product $F_{s}(k,t)S(k,t)$. For a chemically homogeneous system, the following approximate identity holds, $F_{s}(k,t)S(k,t)\sim F(k,t)^{2}$, in the long-time regime.

From an approximate solution to MCT, and also from experiments and simulations, we know that in supercooled liquids $F(k,t)\sim \exp(-t/\tau)^\xi$, with values of the stretching exponent that normally lie in the range $\xi=0.5-0.7$~\cite{Hansen}.
In turn, this argument gives $\nu(t)\sim \exp[-rt^b]$, with stretching exponent $b$ in the range between $0.2$ and $0.3$ for molecular glasses~
\cite{Cui}. 
For metallic glass, we find $b=0.58-0.75$ corresponding to $\xi=0.76-0.85$ which is close to experimental determinations for supercooled metallic melts~\cite{Ruta}, where $\xi\approx 0.8$ in the supercooled regime near the glass transition temperature.

\section{Derivation of Eq.(3) in the main text}

\subsection{Generalized Langevin Equation}
In nonaffine lattice dynamics, Eq.(3) in the main article, without the thermal noise term, is a Generalized Langevin Equation for nonaffine motions in a disordered solid subjected to strain that we derive here for the first time.
Our starting point is Eq.(2) in the main article, which is derived from the Caldeira-Leggett system-bath Hamiltonian in mass-rescaled coordinates:
\begin{equation}
\ddot{q}=-V'(q)-\int_{-\infty}^t \nu(t-t')\frac{dq}{dt'}dt' + F_{p}(t).
\end{equation}
Here, $V(q)$ represents the potential field of the tagged particle at position $q$, and $f=-V'(q)$ represents the force acting on a tagged particle due to its interaction with other particles (atoms) in the material. The thermal noise term $F_{p}(t)$ will be dropped in the subsequent analysis since it has zero mean and it is generally found to vanish upon averaging over several oscillation cycles~\cite{Rodney}.

\subsection{Nonaffine deformations}
In this section we  will use the Eq.(B1) presented in the previous section as a starting point to derive the equation of motion of a tagged atom in a disordered solid metal undergoing an elastic deformation.

Using the same notation as in Ref.~\cite{Lemaitre}, we assume particles lie in a unit cell described by three Bravais vectors $\underline{\underline{h}}=(\underline{a},\underline{b},\underline{c})$. Thus, the interaction potential depends on both $\underline{q}_i$ and $\underline{\underline{h}}$, $\mathcal{U}=\mathcal{U}(\underline{q}_i,\underline{\underline{h}})$ and any vector $\underline{q}$ is mapped onto a cubic reference cell: $\underline{q}=\underline{\underline{h}}\underline{s}, s_{\alpha}\in[-0.5, 0.5]$. We use the unit cell as it is prior to deformation as the reference frame $\underline{\underline{\mathring{h}}}$ and denote the deformed cell by $\underline{\underline{h}}$. When the tagged particle undergoes a displacement to the position $\underline{q}_i$, the process can be understood to consist of two steps: initially, we have  $\underline{q}_i=\underline{\underline{F}}\mathring{\underline{q}}_i$ where $\underline{\underline{F}}=\underline{\underline{h}}\underline{\underline{\mathring{h}}}^{-1}$ is the deformation gradient tensor. $\underline{\underline{F}}$ describes an affine transformation of the unit cell whereas $\mathring{\underline{q}}_i$ remains unchanged. In the second step of the process, particles perform non-affine displacements by relaxing to their nearest equilibrium position $\{\underline{q}_i\}$, keeping $\underline{\underline{h}}$ (and hence $\underline{\underline{F}}$) fixed. Those new coordinates are generally different from the affine positions derived by the reference coordinates, $\{\underline{q}_i\}\neq\{\underline{\underline{F}}\underline{\mathring{q}}_i\}$. For small deformations the non-affine equilibrium positions of the particles are a continuous function of $\underline{\underline{h}}:\{\underline{q}_i\}=\{\underline{q}_i(\underline{\underline{h}})\}$.

\subsection{Deriving the equation of motion for the nonaffine displacement}
We thus rewrite the above Eq.(B1) for a tagged atom in a 3D cell which moves with an affine velocity prescribed by the deformation gradient tensor
$\underline{\underline{\dot{F}}}$:
 \[\underline{\ddot{q}}_i=\underline{f}_i-\int_{-\infty}^t\nu(t-t')\left(\underline{\dot{\mathring{q}}}_i - \underline{u}\right) dt'\]
where $\underline{f}_i=-\partial{\mathcal{U}}/\partial{\underline{q}}_i$ generalizes the $-V'(q)$ in Eq.(B1) to a tagged atom in a 3D lattice.
Furthermore, we used the Galilean transformations to express the particle velocity in the moving frame: $\underline{\dot{q}}_{i}=\underline{\dot{\mathring{q}}}_i - \underline{u}$ where $\underline{u}=\underline{\underline{\dot{F}}}\underline{\mathring{q}}_{i}$ represents the local velocity of the moving frame. This is consistent with our use of the circle on the variables to signify that they are measured with respect to the reference rest frame.

In terms of the original rest frame $\lbrace\underline{\mathring{q}}_i\rbrace$, the equation of motion can be written as,
\begin{equation}
\underline{\underline{F}}\cdot\underline{\ddot{\mathring{q}}}_i  =\underline{f}_i-
\int_{-\infty}^{t}\nu_0e^{-r(t-t')^b}\frac{d\mathring{\underline{q}}_i}{dt'}dt'.
\end{equation}

The terms $\ddot{\underline{\underline{F}}}\underline{\mathring{q}}_i$ and $\int_{-\infty}^{t}\nu_0e^{-r(t-t')^b}\underline{\underline{\dot{F}}}\mathring{\underline{q}}_idt'$ are not allowed into the equation of motion because
they depend on the position of the particle, and therefore have to vanish for a system with translational invariance, as noted already by Andersen~\cite{Andersen} and by Ray and Rahman~\cite{Ray}.

We work in the linear regime of small strain $\parallel\underline{\underline{F}}-\underline{\underline{1}}\parallel\ll1$. We make a perturbative expansion in terms of the small displacement $\{\underline{x}_i(t)=\underline{\mathring{q}}_i(t)-\underline{\mathring{q}}_i\}$ around a known rest frame $\underline{\mathring{q}}_i$.
That is we take: $\underline{\underline{F}}=\underline{\underline{1}}+\delta\underline{\underline{F}}+...$ where $\delta\underline{\underline{F}}=\underline{\underline{\epsilon}}=\underline{\underline{F}}-\underline{\underline{1}}$ is the small parameter, and $\underline{\mathring{r}}_i(t)=\underline{\mathring{r}}_i+\underline{x}_i(t)$ where $\underline{x}_i$ is the nonaffine displacement. We substitute this into equation (B2):

\begin{align}
&(\underline{\underline{1}}+\delta\underline{\underline{F}}+...)\frac{d^2\underline{x}_i}{dt^2}\notag\\
&=\delta\underline{f}_i
-(\underline{\underline{1}}+\delta\underline{\underline{F}}+...)\int_{-\infty}^{t}\nu_0e^{-r(t-t')^b}\frac{d\underline{x}_i}{dt'}dt'.
\end{align}

For the term $\delta\underline{f}_i$, we use the definition of $\underline{\underline{\eta}}$:
\[\underline{\underline{\eta}}=\frac{1}{2}(\underline{\underline{F}}^T\underline{\underline{F}}-\underline{\underline{1}})
\rightarrow\frac{1}{2}(\underline{\underline{\epsilon}}+\underline{\underline{\epsilon}}^T)\]
where the second limit comes from $\|\underline{\underline{F}}-\underline{\underline{1}}\|\ll 1$.
Hence, in this limit of small deformations, $\underline{\underline{\eta}}$ coincides with the strain tensor of linearized elasticity $\underline{\underline{e}}=\frac{1}{2}(\underline{\underline{\epsilon}}+\underline{\underline{\epsilon}}^T)$.
Also considering that $\underline{f}_{i}=0$ identically because of mechanical equilibrium, we have:
\[\delta\underline{f}_i=\frac{\partial\underline{f}_i}{\partial\underline{\mathring{q}}_j} \delta\underline{\mathring{q}}_j
+\frac{\partial\underline{f}_i}{\partial\underline{\underline{\eta}}}:\delta\underline{\underline{\eta}}\]

Hence, upon retaining only zero-order terms in the expansion in Eq.(B3) above, and using the definition of the Hessian
\[\frac{\partial\underline{f}_i}{\partial\underline{\mathring{q}}_j} \delta\underline{\mathring{q}}_j=-\underline{\underline{H}}_{ij}\underline{x}_j\]
and of the nonaffine force:
\[\underline{\Xi}_{i,\kappa\chi}=\frac{\partial\underline{f}_i}
{\partial\eta_{\kappa\chi}}\vert_{\underline{\underline{\eta}}{\rightarrow\underline{\underline{0}}}}\]
we finally obtain for the case of uniaxial elongation:
\begin{equation}
\frac{d^2\underline{x}_i}{dt^2}+\int_{-\infty}^{t}\nu(t-t')\frac{d\underline{x}_i}{dt'}dt'+\underline{\underline{H}}_{ij}\underline{x}_j=\underline{\Xi}_{i,xx}\eta_{xx},
\end{equation}
which is Eq.(3) in the main article.

\section{The Hessian and the affine force-field for binary metallic glasses using the EAM potential}
In order to calculate the dynamics and the viscoelastic response, we need to evaluate the interaction energy between atoms in the material.
In particular, we need to find expressions for the Hessian matrix and for the affine force-field $\underline{\Xi}_{i,\kappa\chi}$, as a function of the interatomic interaction potential. To this aim, we use the embedded-atom model (EAM).
Upon considering the various contributions to the interaction potential between atoms in the CuZr- based MGs, the total potential energy acting on a tagged atom $i$ is (we will drop the label $i$) is given by
\begin{equation}
\mathcal{U}_i=F_{M}(\sum_{j\neq i}\rho_{MN}(Q_{ij}))+\frac{1}{2}\sum_{j\neq i}\psi_{MN}(Q_{ij}).
\end{equation}
Here $q_{ij}$ represents the radial distance of atom $i$ from atom $j$, which is the modulus of the vector $\underline{q}_j-\underline{q}_i$;
$\rho_{N}$ is the contribution to the electronic charge density from atom $j$ of type $N$ at the location of atom $i$ of type $M$; $\psi_{MN}$ is a pairwise potential between an atom of type $M$ and an atom of type $N$, and $F_{M}$ is the embedding function that gives the energy required to place the tagged atom $i$ of type $M$ into the electron cloud. Hence, the total potential is the sum over all particles, $\mathcal{U}=\sum_i\mathcal{U}_i$.

The many-body nature of the EAM potential is a result of the embedding energy term. Both summations in the formula are over all neighbors $j$ of atom $i$ within the cutoff distance~\cite{sutton}. Then we can get the net force acting on a tagged atom using the following set of relations:
\begin{align}
\underline{n}_{ij}&=\frac{\underline{Q}_{ij}}{Q_{ij}}\\
\bar{\rho}_i & =\sum_{j\neq i}\rho_{MN}(Q_{ij}),~\bar{\rho}_j=\sum_{i\neq j}\rho_{MN}(Q_{ij}) \\
Z_{ij}&=\frac{\partial\mathcal{U}_i}{\partial Q_{ij}}=\frac{1}{2}\frac{\partial\psi_{MN}(Q_{ij})}{\partial Q_{ij}}+\frac{\partial F_{M}}{\partial\bar{\rho}_i}\frac{\partial\rho_{MN}(Q_{ij})}{\partial{Q_{ij}}}\\
\underline{f}_i&=-\frac{\partial\mathcal{U}}{\partial\underline{Q}_i}=-\frac{\partial\mathcal{U}_i}{\partial\underline{Q}_i}-\frac{\partial{\sum_{k\neq i}\mathcal{U}_k}}{\partial\underline{Q}_i}\notag\\
&=-\frac{\partial\mathcal{U}_i}{\partial\underline{Q}_i}-\frac{\partial\sum_{k\neq i}\mathcal{U}_k}{\partial{Q_{ik}}}\cdot\frac{\partial Q_{ik}}{\partial\underline{Q}_i}\notag\\
&=-\frac{\partial\mathcal{U}_i}{\partial\underline{Q}_i}+\frac{\partial\sum_{k\neq i}\mathcal{U}_k}{\partial{Q_{ik}}}\frac{ \underline{Q}_{ik}}{Q_{ik}}=-\frac{\partial\mathcal{U}_i}{\partial\underline{Q}_i}+\sum_{k\neq i}Z_{ki}\frac{\underline{Q}_{ik}}{Q_{ik}}.
\end{align}
Note that $\underline{f}_{i}$ and $\underline{\underline{H}}_{ij}$
below, are in general different functions when expressed as functions of bare coordinate $Q$ rather than 
mass-rescaled coordinate $q$, but we use here the same symbols in order to avoid too heavy notation.

The Hessian is then written for $i\neq j$ as:
\begin{align}
\underline{\underline{H}}_{ij}|_{i\neq j}&=\frac{\partial^2\mathcal{U}}{\partial\underline{Q}_i\underline{Q}_j}=\frac{\partial{\frac{\partial\mathcal{U}_i}{\partial\underline{Q}_i}}}{\partial\underline{Q}_j}-\frac{\partial\sum_{k\neq i}Z_{ki}\cdot\frac{\underline{Q}_{ik}}{Q_{ik}}}{\partial\underline{Q}_j}\notag\\
&=\frac{\partial^2\mathcal{U}_i}{\partial\underline{Q}_i\partial\underline{Q}_j}
-\frac{\partial Z_{ji}}{\partial\underline{Q}_j}\cdot\frac{\underline{Q}_{ji}}{Q_{ji}}\notag\\
&-Z_{ji}\frac{\partial\frac{\underline{Q}_{ij}}{Q_{ij}}}{\partial{\underline{Q}_j}}
-\frac{\partial\sum_{k\neq i, k\neq j}Z_{ki}\frac{\underline{Q}_{ik}}{Q_{ik}}}{\partial{\underline{Q}_j}}\notag\\
&=\frac{\partial^2\mathcal{U}_i}{\partial\underline{Q}_i\partial\underline{Q}_j}-\frac{\partial Z_{ji}}{\partial Q_{ij}}\frac{\partial Q_{ij}}{\partial\underline{Q}_j}\otimes\frac{\underline{Q}_{ij}}{Q_{ij}}-Z_{ji}\frac{\partial\frac{\underline{Q}_{ij}}{Q_{ij}}}{\underline{Q}_j}\notag\\
&-\sum_{k\neq i,k\neq j}\frac{\partial Z_{ki}}{\partial\underline{Q}_j}\otimes\frac{\underline{Q}_{ik}}{r_{ik}}
\end{align}

with
\begin{equation}
\frac{\partial\frac{\underline{Q}_{ij}}{Q_{ij}}}{\partial\underline{Q}_j}=\frac{I_{3\times3}}{Q_{ij}}-\frac{\underline{Q}_{ij}\otimes\underline{Q}_{ij}}{Q^3_{ij}},
\end{equation}
and:
\begin{align}
&\underline{\underline{H}}_{ii}=\frac{\partial^2\mathcal{U}}{\partial\underline{Q}_i\underline{Q}_i}\notag\\
&=\frac{\partial^2\mathcal{U}_i}{\partial\underline{Q}_i^2}
-\frac{\partial\sum_{k\neq i}Z_{ki}}{\partial\underline{Q}_j}\frac{\underline{Q}_{ik}}{Q_{ik}}-\sum_{k\neq i}Z_{ki}\frac{\partial\frac{\underline{Q}_{ik}}{Q_{ik}}}{\partial\underline{Q}_i}\notag\\
&=\frac{\partial^2\mathcal{U}_i}{\partial\underline{Q}_i^2}+\frac{\partial\sum_{k\neq i}Z_{ji}}{\partial\underline{Q}_j}\frac{\underline{Q}_{ik}}{Q_{ik}}\otimes\frac{\underline{Q}_{ik}}{Q_{ik}}-\sum_{k\neq i}Z_{ki}\frac{\partial\frac{\underline{Q}_{ik}}{Q_{ik}}}{\partial\underline{Q}_i}\notag\\
&=\frac{\partial^2\mathcal{U}_i}{\partial\underline{Q}_i^2}+\frac{\partial\sum_{k\neq i}Z_{ji}}{\partial\underline{Q}_j}\frac{\underline{Q}_{ik}}{Q_{ik}}\otimes\frac{\underline{Q}_{ik}}{Q_{ik}}+\notag\\
&~~~~+\sum_{k\neq i}Z_{ki}(\frac{I_{3\times3}}{Q_{ik}}-\frac{\underline{Q}_{ik}\otimes\underline{Q}_{ik}}{Q_{ik}^3})
\end{align}
for the diagonal $i=j$ elements.

To find $\underline{\Xi}_{i,\kappa\chi}=\frac{\partial\underline{f}_i}
{\partial\eta_{\kappa\chi}}\vert_{\underline{\underline{\eta}}{\rightarrow\underline{\underline{0}}}}=\sum_j\underline{\Xi}_{ij,\kappa\chi}$, we write

\begin{equation}
\Xi_{ij,\kappa\chi}^{\alpha}=-S_{ij,\alpha\beta}\frac{\partial Q_{ij}^{\beta}}{\partial\eta_{\kappa\chi}}=-\frac{1}{2}S_{ij,\alpha\beta}(\delta_{\beta\kappa}Q^{\chi}_{ij}+\delta_{\beta\chi}Q_{ij}^{\kappa})
\end{equation}

with:
\begin{align}
\underline{\underline{S}}_{ij}&=\frac{\partial^2\mathcal{U}_i}{\partial\underline{Q}_{ij}\partial\underline{Q}_{ij}}\\
&=\frac{\partial}{\partial\underline{Q}_{ij}}\left(\frac{\partial\mathcal{U}}{\partial\underline{Q}_{ij}}\right)\\
&=\frac{\partial}{\partial\underline{Q}_{ij}}\left(\sum_k\frac{\partial\mathcal{U}_k}{\partial\underline{Q}_{ij}}\right)\notag\\
&=\frac{\partial}{\partial\underline{Q}_{ij}}\left(\sum_{k,l\neq k}\frac{\partial\mathcal{U}_k}{\partial Q_{lk}}\frac{\partial Q_{lk}}{\partial\underline{Q}_{ij}}\right)\notag\\
&=\frac{\partial}{\partial\underline{Q}_{ij}}\left(\frac{\partial\mathcal{U}_i}{\partial Q_{ji}}\frac{\partial Q_{ji}}{\partial\underline{Q}_{ji}}+\frac{\partial\mathcal{U}_j}{\partial Q_{ji}}\frac{\partial Q_{ji}}{\partial\underline{Q}_{ji}}\right)\frac{\partial\mathcal{U}_i}{\partial Q_{ji}}\frac{\partial Q_{ji}}{\partial\underline{Q}_{ji}}\notag\\
&=\frac{\partial}{\partial\underline{Q}_{ij}}\left(Z_{ij}\frac{\underline{Q}_{ij}}{Q_{ij}}+Z_{ji}\frac{\underline{Q}_{ij}}{Q_{ij}}\right)\notag\\
&=\frac{\partial}{\partial\underline{Q}_{ij}}\left(Z_{ij}\underline{n}_{ij}+Z_{ji}\underline{n}_{ij}\right)\notag\\
&=\frac{\partial Z_{ij}}{\partial\underline{Q}_{ij}}\underline{n}_{ij}+Z_{ij}\frac{\partial\underline{n}_{ij}}{\partial\underline{Q}_{ij}}+\frac{\partial Z_{ji}}{\partial{\underline{Q}_{ij}}}\underline{n}_{ij}+Z_{ji}\frac{\partial\underline{n}_{ij}}{\partial\underline{Q}_{ji}}\notag\\
&=\frac{\partial}{\partial\underline{Q}_{ij}}\left(\frac{\partial\mathcal{U}_i}{\partial\underline{Q}_{ij}}\right)\underline{n}_{ij}
+Z_{ij}\frac{\partial}{\partial\underline{Q}_{ij}}\left(\frac{\underline{Q}_{ij}}{q_{ij}}\right)\notag\\
&+\frac{\partial}{\partial\underline{Q}_{ij}}\left(\frac{\partial\mathcal{U}_j}{\partial\underline{Q}_{ij}}\right)\underline{n}_{ij}
+Z_{ji}\frac{\partial}{\partial\underline{Q}_{ij}}\left(\frac{\underline{Q}_{ij}}{Q_{ij}}\right)\notag\\
&=\sum_k\frac{\partial\left(\frac{\partial\mathcal{U}_i}{\partial Q_{ij}}\right)}{\partial Q_{ik}}\frac{\partial Q_{ik}}{\partial\underline{Q}_{ij}}\underline{n}_{ij}+Z_{ij}\frac{Q_{ij}-\underline{Q}_{ij}\frac{\partial Q_{ij}}{\partial\underline{Q}_{ij}}}{Q_{ij}^2}\notag\\
&+\sum_k\frac{\partial\left(\frac{\partial\mathcal{U}_j}{\partial Q_{ij}}\right)}{\partial Q_{jk}}\frac{\partial Q_{jk}}{\partial\underline{Q}_{ij}}\underline{n}_{ij}+Z_{ji}\frac{Q_{ij}-\underline{Q}_{ij}\frac{\partial Q_{ij}}{\partial\underline{Q}_{ij}}}{Q_{ij}^2}\notag\\
&=\frac{\partial^2\mathcal{U}_i}{\partial^2Q_{ij}}\underline{n}_{ij}\underline{n}_{ij}+Z_{ij}\frac{(1-\underline{n}_{ij}\underline{n}_{ij})}{Q_{ij}}
+\frac{\partial^2\mathcal{U}_j}{\partial^2Q_{ij}}\underline{n}_{ij}\underline{n}_{ij}\notag\\
&+Z_{ji}\frac{(1-\underline{n}_{ij}\underline{n}_{ij})}{Q_{ij}}
\end{align}

To distinguish $\underline{\underline{S}}$ from $\underline{\underline{H}}$, one can rewrite $\underline{\underline{H}} (i\neq j)$ as
\begin{align}
\underline{\underline{H}}_{ij}&=\frac{\partial^2\mathcal{U}}{\partial\underline{Q}_i\partial\underline{Q}_j}
=\frac{\partial}{\partial\underline{Q}_i}\left(\sum_k\frac{\partial\mathcal{U}_k}{\partial\underline{Q}_j}\right)\\
&=\frac{\partial}{\partial\underline{Q}_i}\left(\sum_{k,l\neq k}\frac{\partial\mathcal{U}_k}{\partial Q_{kl}}\frac{\partial Q_{kl}}{\partial\underline{Q}_j}\right)\notag\\
&=\frac{\partial}{\partial\underline{Q}_i}\left(\sum_{l\neq j}\frac{\partial\mathcal{U}_j}{\partial Q_{jl}}\frac{\partial Q_{jl}}{\partial\underline{Q}_j}+\sum_{k\neq j, l\neq k}\frac{\partial\mathcal{U}_k}{\partial Q_{kl}}\frac{\partial Q_{kl}}{\partial\underline{Q}_{j}}\right)\notag\\
&=\frac{\partial}{\partial\underline{Q}_i}\left(\sum_{l\neq j}\frac{\partial\mathcal{U}_j}{\partial Q_{jl}}\frac{\partial Q_{jl}}{\partial\underline{Q}_j}+\sum_{k\neq j}\frac{\partial\mathcal{U}_k}{\partial Q_{kj}}\frac{\partial Q_{kj}}{\partial\underline{Q}_j}\right)\notag\\
&=\frac{\partial}{\partial\underline{Q}_i}\left(\sum_{l\neq j}\frac{\partial\mathcal{U}_j}{\partial Q_{jl}}\frac{\underline{Q}_{jl}}{\partial Q_{jl}}+\sum_{k\neq j}\frac{\partial\mathcal{U}_k}{\partial Q_{kj}}\frac{\underline{Q}_{jk}}{\partial Q_{jk}}\right)\notag\\
&=\sum_{k\neq j}\frac{\partial}{\partial\underline{Q}_i}\left(\frac{\partial\mathcal{U}_j}{\partial Q_{jk}}\frac{\underline{Q}_{jk}}{Q_{jk}}+\frac{\partial\mathcal{U}_k}{\partial Q_{kj}}\frac{\underline{Q}_{jk}}{Q_{jk}}\right)\notag\\
&=\sum_{k\neq j}\left(\sum_{l\neq j}\frac{\partial}{\partial Q_{jl}}\left(\frac{\partial\mathcal{U}_j}{\partial Q_{jk}}\right)\frac{\partial Q_{jl}}{\partial\underline{Q}_i}\frac{\underline{Q}_{jk}}{Q_{jk}}\right)\\
&+\frac{\partial\mathcal{U}_j}{\partial Q_{ji}}\frac{\partial}{\partial\underline{Q}_i}\left(\frac{\underline{Q}_{ji}}{Q_{ji}}\right)\notag\\
&+\sum_{k\neq j}\left(\sum_{l\neq j}\frac{\partial}{\partial Q_{kl}}\left(\frac{\partial\mathcal{U}_k}{\partial Q_{jk}}\right)\frac{\partial Q_{kl}}{\partial\underline{Q}_i}\frac{\underline{Q}_{jk}}{Q_{jk}}\right)\\
&+\frac{\partial\mathcal{U}_i}{\partial Q_{ji}}\frac{\partial}{\partial\underline{Q}_i}\left(\frac{\underline{Q}_{ji}}{Q_{ji}}\right)\notag\\
&=\sum_{k\neq j}\left(\frac{\partial}{\partial Q_{ji}}\left(\frac{\partial\mathcal{U}_j}{\partial Q_{jk}}\right)\frac{\partial Q_{ji}}{\partial\underline{Q}_i}\frac{\underline{Q}_{jk}}{Q_{jk}}\right)\\
&+Z_{ji}\frac{(-1+\underline{n}_{ij}\underline{n}_{ij})}{Q_{ij}}\notag\\
&+\sum_{k\neq j}\left(\sum_{l\neq k}\frac{\partial}{\partial Q_{kl}}\left(\frac{\partial\mathcal{U}_k}{\partial Q_{kj}}\right)\frac{\partial Q_{kl}}{\partial\underline{Q}_i}\frac{\underline{Q}_{jk}}{Q_{jk}}\right)\\
&+Z_{ij}\frac{(-1+\underline{n}_{ij}\underline{n}_{ij})}{Q_{ij}}\notag\\
&=\sum_{k\neq j}\left(\frac{\partial}{\partial Q_{ji}}\left(\frac{\partial\mathcal{U}_j}{\partial Q_{jk}}\right)\underline{n}_{ij}\underline{n}_{jk}\right)+Z_{ji}\frac{(-1+\underline{n}_{ij}\underline{n}_{ij})}{Q_{ij}}\notag\\
&+\sum_{k\neq j,i}\frac{\partial}{\partial Q_{ki}}\left(\frac{\partial\mathcal{U}_k}{\partial Q_{kj}}\right)\underline{n}_{ik}\underline{n}_{jk}\\
&+\sum_{k\neq i}\frac{\partial^2\mathcal{U}_i}{\partial Q_{ik}\partial Q_{ij}}\underline{n}_{ik}\underline{n}_{ji}\notag\\
&+Z_{ij}\frac{(-1+\underline{n}_{ij}\underline{n}_{ij})}{Q_{ij}}.
\end{align}

Since in the experiment, the sample was stretched along one direction, we let $\kappa=\chi=x$, which gives
\begin{equation}
\Xi_{ij,xx}^{\alpha}=-S_{ij,\alpha x}q^{x}_{ij}.
\end{equation}

\section{Time-frequency conversion and derivation of Eqs.(8)-(9) of the main article}
In this section we present the conversion from viscoelastic response in the time-domain (in which experimental data have been taken) to viscoleastic response in the frequency domain. The converted data have been used for comparison with our theoretical predictions in the main article.

The stress response to a strain $\epsilon(t)$ in the time domain is given by the Boltzmann causality principle as
\begin{equation}
\sigma(t)=\int_{-\infty}^tE(t-t')\dot{\epsilon}(t')dt'
\end{equation}
where $E(t)$ is the time-dependent elastic modulus and $\dot{\epsilon}$ is the strain rate.
We take the Fourier transform of Eq.(D1):
\begin{align}
\tilde{\sigma}(\omega)&=\int_{-\infty}^{\infty}\int_{-\infty}^{\infty}E(t-t')\Theta(t-t')\dot{\epsilon}(t)e^{-i\omega t}dt'dt\notag \\
&=\int_{-\infty}^{\infty}E(u)\Theta(u)e^{-i\omega u}du\int_{-\infty}^{\infty}\dot{\epsilon}(t')e^{-i\omega t}dt
\end{align}
where $\Theta(t)$ is the step function and $u=t-t'$. Note that, the domain of $\sigma(t)$ is generally the whole real line, while the domain of $G(t)$ is defined only for $t>0$.
If the Fourier transform exists, then we can denote it by $\tilde{\sigma}(\omega)$, which is given by
\begin{equation}
\tilde{\sigma}(\omega)=\mathcal{F}[E(t)]\mathcal{F}[\dot{\epsilon}(t)]=\tilde{E}^*(\omega)\tilde{\epsilon}(\omega).
\end{equation}
Note that the second equation is the usual expression of linear stress-strain relation in the frequency domain~\cite{Lemaitre}.\\

In the stress-relaxation experiments presented in the main article, one starts by applying to the (initially relaxed) sample a sudden deformation $\epsilon_0$:\\\\
\begin{align}
\epsilon(t<0)=0 \qquad \epsilon(t>0)=\epsilon_0=\textit{const}.
\end{align}
"Sudden" means that the deformation is applied over a time much shorter than the shortest time-scale of the Maxwell distribution $\tau_{\textit{min}}$, and can thus be modelled as a Heaviside step function. Under these conditions, one can write
\begin{equation}
\dot{\epsilon}_0(t)=\epsilon_0\delta(t),
\end{equation}
where $\delta(t)$ is the Dirac delta-function. From Eq.(D1) and Eq.(D5), one has:
\begin{equation}
\sigma(t)=\int_{-\infty}^t\epsilon_0E(t-t')\delta(t')dt',
\end{equation}
yielding
\begin{align}
\sigma(t<0)=0 \qquad \sigma(t>0)=\epsilon_0E(t).
\end{align}

The experimental data in the time-domain have been fitted with the Kohlrausch empirical function in order to obtain a smooth function for the Fourier transformation.
Also, this allows us to enucleate the $\alpha$-relaxation from the data.
We therefore take the Fourier transform of the empirical Kohlrausch function $\sigma(t)=\sigma_{\infty}+\sigma_0e^{-(t/\tau)^{\beta}}$ used for the fitting of the experimental data which gives:
\begin{equation}
\int_{0}^{\infty}[\sigma_{\infty}+\sigma_0e^{-(t/\tau)^{\beta}}]
e^{-i\omega t}dt=\tilde{E}^*(\omega)\int_0^{\infty}\epsilon_0e^{-i\omega t}dt.
\end{equation}

Upon rearranging terms we thus obtain:
\begin{equation}
\frac{\sigma_{\infty}}{\sigma_0}+i\omega\int_0^{\infty}e^{-(t/\tau)^{\beta}}(\cos{\omega t}-i\sin{\omega t})dt=\tilde{E}^*(\omega)\frac{\epsilon_0}{\sigma_0},
\end{equation}
by using $\int_0^{\infty}e^{-i\omega t}dt=\pi(1+i)\delta(\omega)+\frac{1}{i\omega}$.
This simplifies to the real and imaginary part of $\tilde{E}^*(\omega)=E'(\omega)+iE''(\omega)$, which are Eq. (8) and Eq.(9), respectively, in the main text.

\end{document}